\documentclass{aa}

\usepackage{ifthen}
\usepackage{graphicx}
\usepackage{natbib}

\bibpunct{(}{)}{;}{a}{}{,}

\newcommand{\pol}{\mbox{Polaris}}
\newcommand{\Lft}{\mbox{L1512}}

\renewcommand{\r} {\vec{r}}
\newcommand{\vlag}{\vec{l}}
\newcommand{\slag}{\emm{l}}

\newcommand{\sdev}[1][]{\emm{\sigma_{#1}}}
\newcommand{\Centroid}[1][]{\emm{C_\emr{#1}}}
\newcommand{\Cincr}[1][]{\emm{\delta \Centroid[#1]}}
\newcommand{\AveCincr}[1][]{\emm{\abs{\Cincr[#1]}}}
\newcommand{\Ucentroid}{\emm{\Delta \Centroid}}

\newcommand{\ObsGrad}{\emm{\nabla v_\emr{obs}}}
\newcommand{\TurbGrad}{\emm{\nabla v_\emr{turb}}}

\newcommand{\Ntot}{\emm{\overline{N}}}
\newcommand{\SNR}[1][]{\mbox{SNR$_{#1}$}} 
\newcommand{\WindowWidth}[1][\mbox{}]{\emm{W_{#1}}}

\newcommand{\Nsign}{\emm{N_\emr{sign}}} 
\newcommand{\xsize}{\emm{N}}
\newcommand{\ysize}{\emm{M}}

\newcommand{\ie} {{\em i.e.}}
\newcommand{\eg} {{\em e.g.}}

\newcommand{\emm}[1]{\ensuremath{#1}}       
\newcommand{\emr}[1]{\emm{\mathrm{#1}}}     
\newcommand{\nds}[1]{\emm{\displaystyle#1}} 

\newcommand{\abs}[1]    {\emm{\left| #1 \right|}}   
\newcommand{\bracket}[1]{\nds{\left[  #1 \right] }} 

\newcommand{\about}     {\emm{\sim}}                     
\newcommand{\Definition}{\emm{\equiv}}                   

\newcommand{\HI}  {\ion{H}{i}}       
\newcommand{\HH}  {\mbox{H$_2$}}     
\newcommand{\thCO}{\mbox{$^{13}$CO}} 
\newcommand{\twCO}{\mbox{$^{12}$CO}} 
\newcommand{\CeiO}{\mbox{C$^{18}$O}} 
\newcommand{\HCOp}{\mbox{HCO$^{+}$}} 
\newcommand{\Jone}{\mbox{(J=1--0)}}
\newcommand{\Jtwo}{\mbox{(J=2--1)}}

\newcommand{\unit}[1]{\emm{\, \emr{#1}}}
\newcommand{\K}{\unit{K}}
\newcommand{\m}{\unit{m}}
\newcommand{\au}{\unit{AU}}
\newcommand{\pc}{\unit{pc}}
\newcommand{\kms}{\unit{km\,s^{-1}}}
\newcommand{\Kkms}{\unit{K\,km\,s^{-1}}}
\newcommand{\pscm}{\unit{cm^{-2}}}
\newcommand{\pccm}{\unit{cm^{-3}}}

\newcommand{\FigureSignalPolaris}{%
\begin{figure}
  \centering \resizebox{\hsize}{!}{\includegraphics{ms2555.f1}}
  \caption{Maps of the Polaris (MCLD123+24.9) field
    in the \twCO\Jone{} {\it (left)} and \thCO\Jone{} {\it (right)} lines:
    {\it (top)} line integrated areas, {\it (bottom)} line centroid
    velocities. The offsets are in arcsec respectively to the (0,0)
    position: $l_{II}=123.68^\circ$, $b_{II}=24.93^\circ$.  The two
    upper-left corners of the \twCO{} and \thCO{} maps coincide. The dashed
    contour level at 4\Kkms{} in the $W_{\thCO}$ map is meant to localize
    the region of largest column density and is reproduced in the
    $W_{\twCO}$ map. The black contour is the 0.6 \Kkms{}--level of the
    CS(3-2) map (see Fig.~\ref{fig:dc}) and is meant to localize the dense
    core in the rest of the paper. The two crosses show the positions of
    the HC$_3$N peaks~\citep{heithausen02:gbcmcc}. The linear scale is
    shown.}
  \label{fig:sig:pol}
\end{figure}}

\newcommand{\FigureSignalLFifteenTwelveIRAM}{%
\begin{figure}
  \centering \resizebox{\hsize}{!}{\includegraphics{ms2555.f2}}
  \caption{Same as Fig.~\ref{fig:sig:pol} for the L1512 field observed at
    IRAM-30m. Offsets are in arcsec relative to the (0,0) position:
    RA(1950)= 05h00m54.5s, Dec(1950)=32$^\circ$39'00''.  The dashed contour
    level (3\Kkms{} in the $W_{\thCO}$ map) is reproduced in the
    $W_{\twCO}$ map and helps localize the regions of largest column
    density. The dense core here (black contour) is identified with the
    0.15\Kkms{} contour of the \HCOp{}(3-2) emission (see
    Fig.~\ref{fig:dc}) and is reproduced in the rest of the figures. The
    linear scale is shown.}
  \label{fig:sig:l1512}
\end{figure}}

\newcommand{\FigureSignalLFifteenTwelveCSO}{%
\begin{figure}
  \centering \resizebox{\hsize}{!}{\includegraphics{ms2555.f3}}
  \caption{Maps of the \twCO\Jtwo{} line integrated area $W$
    {\it (top)} and line centroid velocities $C$ {\it (bottom)} of the
    L1512 field observed at the CSO. In both panels, a map of the same
    quantities computed with the high angular resolution data of the
    IRAM-30m is inserted in the CSO map at the relevant position. The dense
    core is identified as in Fig.~\ref{fig:dc} (black contour).  The dashed
    contour corresponds to $W_{\twCO}=2.5 $\Kkms{} and will be used in
    Fig.~\ref{fig:cvi:l1512s}.  The box is the location of the spectra map
    shown in Fig.~\ref{fig:smp:l1512}.  The (0,0) position of the CSO map
    is RA(1950)= 05h00m54.5s, Dec(1950)=32$^\circ$36'00''. The linear scale
    is shown.}
  \label{fig:sig:l1512s}
\end{figure}}

\newcommand{\FigureDenseCores}{%
\begin{figure}
  \centering \resizebox{0.75\hsize}{!}{\includegraphics{ms2555.f4}}
  \caption{Map of integrated emission of the dense cores: IRAM-30m map
    of Polaris in the CS(3-2) line {\it(top)} and CSO map of L1512 in the
    \HCOp(3-2) line {\it (bottom)}. }
  \label{fig:dc}
\end{figure}}

\newcommand{\FigureSpectraMap}{%
\begin{figure}
  \centering \resizebox{\hsize}{!}{\includegraphics{ms2555.f5}}
  \caption{Representative spectra of the \twCO\Jtwo{} CSO map, located
    in the box drawn in Fig.~\ref{fig:sig:l1512s}. On each spectrum, the
    window selected by the procedure designed to optimize the
    signal--to--noise ratio of the line integrated area is shown as thin
    vertical bars, as is the computed line centroid. The line temperature
    and velocity scales are given in the lower left corner.}
  \label{fig:smp:l1512}
\end{figure}}

\newcommand{\FigurePDFsPolaris}{%
\begin{figure}
  \centering \resizebox{\hsize}{!}{\includegraphics{ms2555.f6}}
  \caption{PDFs of centroid velocity increments for increasing lag
    values $l$ given in arcsec and pixels within each frame.  The PDFs have
    been computed from \emph{a)} the \twCO{} \Jone{} and \emph{b)} \thCO{}
    \Jone{} spectra maps of \pol{}. The centroid increment scales have been
    normalized so that $\langle \Cincr{} \rangle = 0$ and $\sdev[\Cincr{}]
    = 1$.  The actual velocity dispersion of each distribution
    $\sigma_{\Cincr(l)}$ is given in \kms\ in each frame.  The dotted
    curves represent a Gaussian of zero mean and unit standard deviation.}
  \label{fig:PDFs:pol}
\end{figure}}

\newcommand{\FigurePDFsLFifteenTwelve}{%
\begin{figure}
  \centering \resizebox{\hsize}{!}{\includegraphics{ms2555.f7}}
  \caption{Same as Fig.~\ref{fig:PDFs:pol} for the L1512 field. The PDFs
    have been computed from the \textit{a)} \twCO\Jone\ and \textit{b)}
    \thCO\Jone{} IRAM-30m maps and \textit{c)} \twCO{} \Jtwo{} CSO map.}
  \label{fig:PDFs:l1512}
\end{figure}}

\newcommand{\FigureSDEVvsLag}{%
\begin{figure}
  \centering \resizebox{\hsize}{!}{\includegraphics{ms2555.f8}}
  \caption{Scaling of the standard deviations of centroid increments
    with the lag $l$ used to compute the increments for the five fields and
    lines. Large symbols are used for the lags which give statistically
    meaningful results as explained in Appendix~\ref{sec:sampling}.  The
    exponents $\zeta_2$ listed in Table~\ref{tab:lvg} have been determined
    for the range of lags which is statistically meaningful.}
  \label{fig:sdev-vs-lag}
\end{figure}}

\newcommand{\FigureCVIMapPolaris}{%
\begin{figure}
  \centering \resizebox{\hsize}{!}{\includegraphics{ms2555.f9}}
  \caption{Top panels: Maps of the averaged centroid increments 
    \AveCincr{} in Polaris computed with a lag of 3 pixels from the
    \twCO\Jone{} lines {\it (left)} and the \thCO\Jone{} lines {\it
      (right)}.  The solid contour localizes the dense core (see
    Fig.~\ref{fig:sig:pol}). The bottom panels are sketches meant to
    emphasize the relative orientation of the region of large column
    density traced by \thCO\ (grey area), the dense core (dashed contour)
    and regions of large centroid increments where \AveCincr{} $>$
    \AveCincr[0] (black contours).  The thresholds \AveCincr[0] are
    those listed in Table~\ref{tab:cvi-stat}.  The letters are the labels
    used in the text.}
  \label{fig:cvi:pol}
\end{figure}}

\newcommand{\FigureCVIMapLFifteenTwelveIRAM}{%
\begin{figure}
  \centering \resizebox{\hsize}{!}{\includegraphics{ms2555.f10}}
  \caption{Same as Fig.~\ref{fig:cvi:pol} for the L1512 field observed at 
    IRAM. }
  \label{fig:cvi:l1512}
\end{figure}}

\newcommand{\FigureCVIMapLFifteenTwelveCSO}{%
\begin{figure}
  \centering \resizebox{\hsize}{!}{\includegraphics{ms2555.f11}}
  \caption{Same as Fig.~\ref{fig:cvi:pol} for the large scale L1512
    field. The dot-dashed contour is the same as in
    Fig.~\ref{fig:sig:l1512s}.}
  \label{fig:cvi:l1512s}
\end{figure}}

\newcommand{\FigureCVICutPolarisDec}{%
\begin{figure}
  \centering \resizebox{\hsize}{!}{\includegraphics{ms2555.f12}}
  \caption{Cuts in the \twCO\ line of the Polaris field in RA, averaged
    over the indicated Dec range. The histograms are the cuts of increment
    centroids: raw values in the top panel, averaged in azimuth as
    explained in Section~\ref{sec:spatial:not-random} in the bottom panel.
    The other cuts are the centroid velocities \emph{(top)} and integrated
    area \emph{(bottom)}. The letters correspond to those of
    Fig.~\ref{fig:cvi:pol}.}
  \label{fig:cut:pol:dec}
\end{figure}}

\newcommand{\FigureCVICutPolarisRA}{%
\begin{figure}
  \centering \resizebox{\hsize}{!}{\includegraphics{ms2555.f13}}
  \caption{Same as Fig.~\ref{fig:cut:pol:dec} for the \thCO\ line of the 
    Polaris field. Here the cuts are in RA, averaged over the indicated Dec
    range.}
  \label{fig:cut:pol:ra}
\end{figure}}

\newcommand{\FigureCVICutLFifteenTwelveCSODec}{%
\begin{figure}
  \centering \resizebox{\hsize}{!}{\includegraphics{ms2555.f14}}
  \caption{Same as Fig.~\ref{fig:cut:pol:dec} for the large scale L1512 
    field. The letters correspond to those of Fig.~\ref{fig:cvi:l1512s}.
    \vspace*{1.55cm}}
  \label{fig:cut:l1512s:dec}
\end{figure}}

\newcommand{\FigureCVICutLFifteenTwelveCSORA}{%
\begin{figure}
  \centering \resizebox{\hsize}{!}{\includegraphics{ms2555.f15}}
  \caption{Same as Fig.~\ref{fig:cut:pol:ra} for the large scale L1512 
    field.}
  \label{fig:cut:l1512s:ra}
\end{figure}}

\newcommand{\FigureZoomLFifteenTwelve}{%
\begin{figure}
  \centering \resizebox{\hsize}{!}{\includegraphics{ms2555.f16}}
  \caption{Maps of average centroid increments of the same region of the
    L1512 field computed from the IRAM-30m \twCO\Jone{} map {\it (left)} 
     and the CSO \twCO\Jtwo{} map {\it (right)}. Note the
    remarkable quantitative agreement between the two maps.}
  \label{fig:zoom:l1512}
\end{figure}}

\newcommand{\FigureWCOvsCVI}{%
\begin{figure}
  \centering \resizebox{\hsize}{!}{\includegraphics{ms2555.f17}}
  \caption{Scatter plots of the integrated emission versus the
    centroid increments for three data sets analyzed.}
  \label{fig:wco-vs-cvi}
\end{figure}}

\newcommand{\FigureRAIvsCVI}{%
\begin{figure}
  \centering \resizebox{\hsize}{!}{\includegraphics{ms2555.f18}}
  \caption{Scatter plots of the azimuthally averaged
    increments, \AveCincr{}, versus the relative variations of the line
    integrated area, \emm{\abs{\delta A}}/A, for three data sets.  The
    horizontal lines show the \AveCincr[0] thresholds separating the
    Gaussian kernel from the non-Gaussian wings in the \Cincr{}-PDFs. The
    vertical lines are an estimation of the required column density
    relative variations to produce centroid increments of the order of
    \AveCincr[max] in each field.}
  \label{fig:rai-vs-cvi}
\end{figure}}

\newcommand{\FigureSNRvsCVI}{%
\begin{figure}
  \centering \resizebox{8cm}{!}{\includegraphics{ms2555.fA1}}
  \caption{Scatter plots of the \Cincr{} signal--to-noise ratio (defined as
    the ratio of \Cincr{} to its uncertainty) versus the \Cincr{}-values for
    the five data sets analyzed. These scatter plots show that the
    structures delineated by the largest centroid increments and shown in
    Figs.~\ref{fig:cvi:pol} to~\ref{fig:cvi:l1512s} have large SNRs.}
  \label{fig:snr-vs-cvi}
\end{figure}}

\newcommand{\FigureNormalizedPDFsShapeVSNoise}{%
\begin{figure}
  \centering \resizebox{\hsize}{!}{\includegraphics{ms2555.fA2}}
  \caption{ Results of numerical simulations showing the evolution
    of the shape of the normalized \Cincr{}--PDFs as thermal noise is
    progressively added to maps of noiseless Gaussian line profiles.  The
    two maps have the same spatial distribution of line areas and widths as
    the real data of the L1512 large scale field.  Their difference lies in
    the distribution of line centroids: it is Gaussian (top panels) or
    identical to the real data (bottom panels) (see text). Two different
    \SNR[a] thresholds (i.e. 3 and 10) are used in the computation of the
    centroid velocity increments. For the Gaussian velocity field, thermal
    noise can create non-Gaussian wings to the \Cincr{}--PDFs, only when the
    \SNR[a] threshold is 3 and for the largest noise. In comparison, the
    evolution in the bottom panels shows that the L1512 velocity field has
    intrinsic non-Gaussian statistical properties. The Gaussian PDF in the
    case of largest noise and \SNR[a] is due to poor sampling (see text).}
  \label{fig:PDFs:norm}
\end{figure}}

\newcommand{\FigureNumberOfSpectraVSNoise}{%
\begin{figure}
  \centering \resizebox{6cm}{!}{\includegraphics{ms2555.fA3}}
  \caption{Evolution of the number of spectra whose \SNR[a] value is above
    a given threshold (3 for the dotted line and 10 for the plain line) as
    a function of the noise level.}
  \label{fig:npts-vs-noise}
\end{figure}}

\newcommand{\TableTelDescription}{%
  \begin{table*}
    \begin{center}
      \caption{Characteristics of the maps: telescope,
        line observed, half-power beamwidth, sampling step (or pixel size)
        in arc sec and in AU at the distance of the sources (150\pc{}), the
        effective resolution in pixels and in AU}
      \label{tab:tels}
      \begin{tabular}{ccccc}
        \hline
        Telescope & Line & HPBW & Pixel Size & Resolution \\
        \hline
        IRAM (30\m) & CO \Jone
        & 22" & 7.5" (1\,125\au) & 3 pixels (3\,375\au) \\
        CSO  (10.4\m) & CO \Jtwo
        & 28" &  16" (2\,400\au) & 2 pixels (4\,800\au) \\
        \hline
      \end{tabular}
    \end{center}
  \end{table*}}

\newcommand{\TableFieldDescription}{%
  \begin{table*}
    \begin{center}
      \caption{Characteristics of the mapped fields: sizes in \pc,  
        number of spectra and average signal-to-noise ratios in peak
        temperature and integrated area (The associated uncertainties are
        the rms scatter of the values over the maps).}
      \label{tab:fields}
      \begin{tabular}{cccccc}
        \hline
        Source & Line & Field size (pc) & Nb. Spectra & SNR & SNR$_a$ \\
        \hline
        \pol & \twCO{} \Jone & 0.26 $\times$ 0.35 & 3300 & 11$\pm$ 3 & 51$\pm$ 13\\
        \pol & \thCO{} \Jone & 0.22 $\times$ 0.30 & 1650 & 12$\pm$ 4 & 40$\pm$ 12\\
        \Lft & \twCO{} \Jone & 0.22 $\times$ 0.44 & 3200 & 15$\pm$ 3 & 46$\pm$ 8 \\
        \Lft & \thCO{} \Jone & 0.22 $\times$ 0.36 & 2560 & 23$\pm$ 5 & 61$\pm$ 13\\
        \Lft & \twCO{} \Jtwo & 1.05 $\times$ 1.12 & 8300 & 10$\pm$ 4 & 27$\pm$ 12\\
        \hline
      \end{tabular}
    \end{center}
  \end{table*}}

\newcommand{\TableNormalization}{%
  \begin{table}
    \begin{center}
      \caption{Quantities used to normalize the \Cincr{}--PDFs for each field
        and line and for the lag $l=3$ pixels: the offset applied to the
        increments to get a centered PDF (col.~3) and the standard
        deviation of the \Cincr{} values (col.~4). The ratio of these two
        quantities is given in col.~5.}%
      \label{tab:norm}
      \medskip
      \begin{tabular}{ccccc} 
        \hline 
        Field & Line & $\langle \Cincr[3] \rangle$ & $\sigma_{\Cincr[3]}$
        &${\langle \Cincr[3] \rangle \over \sigma_{\Cincr[3]}}$ \\
        & & \kms &\kms & \\
        \hline 
        Polaris &\twCO{} \Jone & 8.3 $\times 10^{-3}$& 0.11 &0.07 \\
        Polaris &\thCO{} \Jone &1.5 $\times 10^{-2}$ & 0.10 &0.15 \\
        L1512 & \twCO{} \Jone &1.2 $\times 10^{-2}$ &0.05 &0.24 \\
        L1512 & \thCO{} \Jone &1.8 $\times 10^{-3}$ &0.04 &0.04 \\
        L1512 & \twCO{} \Jtwo{} &1.3 $\times 10^{-2}$ &0.09 & 0.015 \\
        \hline
      \end{tabular} 
    \end{center}
  \end{table}}

\newcommand{\TableLargeVelocityGradient}{%
  \begin{table*}
    \begin{center}
      \caption{Influence of large scale velocity gradients. Difference
        between the observed extrema of line centroids in each map
        (col.~3), the associated scale $s$ (col.~4) and the corresponding
        observed large scale velocity gradients (col.~5), the contribution
        of this large scale gradient to a lag $l=3$ pixels (col.~6), the
        ratio of this contribution to the internal velocity dispersion at
        the same scale (col.~7), the exponent of the scaling of the
        standard deviation of the centroid increments with the lag at which
        they are computed (col.~8), the velocity gradients at scale $s$
        inferred from this scaling and ascribed to turbulence (col.~9).}
      \label{tab:lvg}
      \medskip
      \begin{tabular}{ccccccccc}
        \hline
        Field & Line & \abs{\Centroid[min]-\Centroid[max]} & $s$ & \ObsGrad{}
        &$\delta v_{3}$&${\delta v_{3} \over \sigma_{\Cincr[3]}}$  & $\zeta_2$ & \TurbGrad{}\\
        &   &  \kms & "/pc &\kms{} pc$^{-1}$& \kms &  &  &\kms{} pc$^{-1}$ \\
        \hline
        Polaris & \twCO(1-0) & 1.1 & 500/0.37 & 3.0 & 0.04 & 0.35 & 0.45 & 3.8\\
        Polaris & \thCO(1-0) & 1.0 & 400/0.29 & 3.4 & 0.06 & 0.60 & 0.5  & 4.5\\
        L1512   & \twCO(1-0) & 1.2 & 600/0.44 & 2.7 & 0.05 & 1.0  & 0.62 & 2.7\\
        L1512   & \thCO(1-0) & 0.4 & 300/0.22 & 1.8 & 0.03 & 0.77 & 0.7  & 3.5\\
        L1512   & \twCO(2-1) & 1.4 & 1200/0.9 & 1.5 & 0.03 & 0.33 & 0.5  & 1.6\\
        \hline
      \end{tabular}
    \end{center}
  \end{table*}}

\newcommand{\TableCVIsStatistics}{%
  \begin{table*}
    \begin{center}
      \caption{Characteristics of the maps of averaged centroid
        increments for each field and line: thresholds used to define the
        subsets of positions populating the non-Gaussian wings of
        \Cincr{}--PDFs (col. 3), peak increment (col. 4) and background (col.
        5) values, square of the contrast of the peak to background value
        (col. 6), average of the line profile velocity dispersion over the
        field (col. 7), surface filling factor of regions where
        $\AveCincr{}>\AveCincr[0]$ (col. 8) and fraction of
        dissipation in these regions (col. 9, see Section 7). The last
        column is the average linewidth in each field (see Section 6.3).  }
      \label{tab:cvi-stat}
      \medskip
      \begin{tabular}{cccccccccc}
        \hline
        Field & Line & \AveCincr[0] & \AveCincr[max] & 
        \AveCincr[bg] & $(\Cincr[max]/\Cincr[bg])^2$ &
        $\langle \sigma_u \rangle$ & $f_s$ & $f_{\epsilon}$ & $\langle
        \Delta v \rangle$ \\
        &     &  \kms  &\kms        & \kms & & \kms  & &  & \kms\\
        \hline
        Polaris & \twCO{} \Jone & 0.1  & 0.4  & 0.06 & 39 &0.8& 0.29 &
        0.65 & 0.8\\
        Polaris & \thCO{} \Jone & 0.1  & 0.4  & 0.05 & 82 &0.4& 0.20 &
        0.62& 0.38\\
        L1512   & \twCO{} \Jone & 0.05 & 0.1  & 0.03 & 15 &0.3& 0.29 &
        0.61& 0.30\\
        L1512   & \thCO{} \Jone & 0.03 & 0.08 & 0.02 & 17 &0.2& 0.39 &
        0.74& 0.17\\
        L1512   & \twCO{} \Jtwo & 0.09 & 0.4  & 0.06 & 45 & 0.3&0.18 &
        0.53& 0.29\\
      \hline
    \end{tabular}
  \end{center}
\end{table*}}

\newcommand{\Tablelag}{%
  \begin{table}
    \begin{center}
      \caption{Largest significant lags for the \Cincr{}--PDFs in each
        field. The field sizes are \xsize\ and \ysize\ in pixels and the
        largest significant lags are given (in pixels) for \Cincr{}--PDFs
        built with 8 (col. 6) and 30 (col. 7) bins.}
      \label{tab:lags}
      \begin{tabular}{cccccc}
        \hline
        Source & Line  & \xsize & \ysize & $l_\emr{max}(8)$ & $l_\emr{max}(30)$ \\
        \hline
        \pol & \twCO{} \Jone & 48 & 64 & 12 & 6 \\
        \pol & \thCO{} \Jone & 40 & 56 & 12 & 6 \\
        \Lft & \twCO{} \Jone & 40 & 80 & 13 & 7 \\
        \Lft & \thCO{} \Jone & 56 & 40 & 12 & 6 \\
        \Lft & \twCO{} \Jtwo & 90 & 96 & 21 & 11\\
        \hline
      \end{tabular}
    \end{center}
  \end{table}}

\begin{document}

\title{Non-Gaussian velocity shears \\
  in the environment of low mass dense cores}

\author{J. Pety\inst{1,2} \and E. Falgarone\inst{1}}

\offprints{J. Pety, \email{pety@iram.fr}}

\institute{%
  LERMA/LRA, Observatoire de Paris \& Ecole Normale Sup\'erieure, %
  24 rue Lhomond, F-75005 Paris, France %
  \and Institut de Radio Astronomie Millim\'etrique, 300 rue de la Piscine,
  F-38406 Saint Martin d'H\`eres \\
  \email{pety@iram.fr, falgarone@lra.ens.fr}}

\date{Received xxxx / Accepted xxxx}

\titlerunning{Structures of large velocity shear around dense cores}

\authorrunning{Pety \& Falgarone}

\abstract{%
  We report on a novel kind of small scale structure in molecular clouds
  found in IRAM-30m and CSO maps of \twCO{} and \thCO{} lines around low
  mass starless dense cores.  These structures come to light as the locus
  of the extrema of velocity shears in the maps, computed as the increments
  at small scale ($\sim 0.02$ pc) of the line velocity centroids. These
  extrema populate the non-Gaussian wings of the shear probability
  distribution function (shear-PDF) built for each map. They form elongated
  structures of variable thickness, ranging from less than 0.02 pc for
  those unresolved, up to 0.08 pc.  They are essentially pure velocity
  structures.  We propose that these small scale structures of velocity
  shear extrema trace the locations of enhanced dissipation in interstellar
  turbulence.  In this picture, we find that a significant fraction of the
  turbulent energy present in the field would be dissipating in structures
  filling less than a few \% of the cloud volume.  \keywords{ISM: evolution
    - ISM: kinematics and dynamics - ISM: molecules - ISM: structure -
    Turbulence}}

\maketitle %

\section{Introduction}
\label{sec:intro}

Star formation proceeds at vastly different rates, in space and time,
within a given galaxy and from one galaxy to another. These rates are known
to be governed by the local conditions prevailing in dense and cold gas,
but also depend on large scale environments, up to extragalactic scales.
The only two processes, together with rotation, able to mitigate the
effects of gravity are magnetic fields and supersonic turbulence because
they involve energies of the same order of magnitude as the gravitational
energy in the densest phases of the interstellar medium (ISM). More
precisely, turbulent energy, because of its steep power spectrum, can
stabilize the largest masses, first prone to gravitational instability, a
property not shared by thermal energy which is scale--free
~\citep{panis98:nsevpptf}. On the other hand, it has also been proposed
that supersonic turbulence might trigger star formation in
shocks~\citep{klessen00:gctmc:gt}. The possible role of turbulence in the
star formation process is the motivation behind the plethora of recent
studies on interstellar turbulence.

Important insights to the field have been provided by the determination of
the multi--scale properties of interstellar clouds and their comparison to
the scaling laws of turbulence.  A broad variety of statistical tools have
been used: wavelet analysis ~\citep{gill90:fuwamc,langer93:hsaicunow},
$\Delta$--variance \citep{bensch01:qmcsudv}, auto-correlation
function~\citep{kleiner84:lsstmc:df-fjl, kleiner85:lsstmc:avft,
  dickman85:lsstmc:mt, perault86:fmc:saed}, structure
functions~\citep{miesch94:satmc, miesch95:etcvdsfr, miesch99:vfssfr:cvo,
  padoan03:sfstpmcc}, analysis in terms of fractal
structures~\citep{bazell88:fsic, falgarone91:femc:fbds, stutzki98:fsmc}.
The principal component analysis has been used to diagnose the large--scale
flows of atomic gas into which turbulence in molecular clouds is
embedded~\citep{brunt03:utmimede}.  Observations of dense cores and their
environment have revealed a break of the scaling properties of molecular
clouds at the scale of the dense cores \citep{falgarone98:ikp:ssspsfr,
  goodman98:cdc:tc}.  Direct numerical simulations of compressible
turbulence have also been used to compare the real observables of the
interstellar clouds to those simulated.  The first attempt
by~\citet{falgarone94:sstc}, using the high resolution simulations of midly
compressible turbulence of~\citet{porter94:ksdtdsf} has been followed by
more sophisticated comparisons such as those based on the spectral
correlation function method~\citep{rosolowsky99:scf:ntaslm}.  Numerical
simulations of turbulence including magnetic fields,
e.g.~\citet{ostriker01:dvmfstmcm}, and
self-gravity~\citep{klessen00:gctmc:gt,heitsch01:tctmc:mhdt} provided
further support to the fact that interstellar turbulence bears many of the
statistical properties of supersonic magneto--hydrodynamical (MHD)
turbulence, as simulated.
 
One major surprise brought to the field by direct numerical simulations of
MHD turbulence has been that magnetic fields do not delay the dissipation
of supersonic turbulence~\citep{maclow98:kedrssatsfc}.  This unexpected
result calls for further numerical and observational approaches.  The
present paper is observational: It is an attempt to disclose kinematic
signatures of turbulent dissipation in molecular clouds.

To discuss what these kinematic signatures may be, we first need to briefly
recall the main drivers of dissipation in interstellar turbulence, assuming
infinite conductivity and thus neglecting Ohmic dissipation.  The primary
source of dissipation of supersonic turbulence is shocks, but shock
interaction generates vorticity, as do non--planar shocks
~\citep{porter94:ksdtdsf}. Since the divergence of the velocity field
eludes direct detection in space, shocks cannot be detected by this
kinematic signature, and remanent vorticity is a plausible signature of
fossil shocks.  Viscous dissipation is also present and is due primarily to
elastic collisions. Dissipation due to neutral-neutral collisions follows
the shear of the velocity field, and therefore the
vorticity~\citep{landau87:fm}, and that due to ion-neutral collisions
increases with the drift velocity of the ions relative to the
neutrals~\citep{kulsrud69:ewpipcr}.  This drift causes a force on the ions
which, over timescales longer than the ion-neutral collision
time~\citep{zweibel88:adddtg}, is balanced by the Lorentz force ${\bf
  (\nabla \times B) \times B}$.  The dissipation driven by ion-neutral
collisions therefore involves ${\bf J=\nabla \times B}$, the current
density.

Now, both vorticity, in hydrodynamical turbulence, and current density, in
plasma turbulence, are known to be intermittent in space and time.
Laboratory experiments of incompressible turbulence show the formation of
transient long and thin coherent vortices at the edge of which the velocity
shear is so large that a significant fraction of the viscous dissipation
occurs there~\citep{douady91:doiivft}.  In plasma turbulence, intermittency
also exists and is at the origin of non-Gaussian probability distribution
functions (PDFs) of current density.  It is seen in Tokamak plasma
turbulence~\citep{wang99}.  Observations in the solar wind reveal an
intermittent dissipation as well. The intermittency of dissipation of
plasma turbulence has been modelled by~\citet{politano95:mimhdt}, who
propose that the dissipative structures are sheetlike structures of intense
current density, in agreement with recent 3-dimensional numerical
simulations of MHD turbulence~\citep{politano95,biskamp00}.  A recent
attempt at estimating the sizescales of dissipation in incompressible MHD
turbulence~\citep{cho02:nrmhdt:cbvc} shows that magnetic structures develop
at scales much smaller than the viscous damping scale.

Turbulence in molecular clouds is neither laboratory turbulence nor plasma
turbulence, but the above elements suggest that dissipation follows the
vorticity and the current density, and that these quantities are
intermittent, \ie{} exhibit large fluctuations at small scales. The origin
of currents in molecular clouds is not known, but is likely associated to
differential rotation within clouds at the origin of toroidal or helical
fields~\citep{joulain98:necdsit}. Such helical fields have been inferred
from various observations of molecular clouds~\citep{Hanawa93:emfrffmc,
  joulain98:necdsit, carlqvist98:hsret, falgarone01:fshmfesdc,
  matthews02:mfsfmc:iii, matthews01:mfsfmc:ii}.  These fields have been
invoked to explain the polarization patterns of the dust continuum emission
of filaments of matter~\citep{harjunpaa99:lpmfc, fiege00:psefmc} and their
gravitational stability~\citep{fiege00:hffmc:i, fiege00:hffmc:ii}.
Enhanced vorticity in molecular clouds should therefore be a good tracer of
several major dissipative processes of turbulence in weakly ionized
molecular clouds.  Its kinematic signature, if available in spectral
observations of molecular clouds, is essential to search for because it may
trace fossil shocks, regions of large viscous dissipation in the neutrals,
or local gas differential rotation at the origin of currents.

In this paper, we analyze the velocity structure of several fields in
molecular clouds in order to locate and characterize the regions of
enhanced vorticity, likely to be regions of enhanced dissipation. We focus
on the environment of low mass dense cores almost thermally supported,
where dissipation is anticipated to occur, or has occurred in a recent
past.  Section~\ref{sec:method} gives the method used to trace vorticity.
Section~\ref{sec:fields} is a description of the target fields.  The
statistical analysis of the velocity fields, and the manifestations of
their non-Gaussian features are given in Section~\ref{sec:stat}. The
spatial distribution of these non-Gaussian events is compared to that of
the dense gas in Section~\ref{sec:spatial}. The possible biases of the
method are discussed in Section~\ref{sec:discussion}. The implications of
this study in terms of turbulence dissipation are discussed in
Section~\ref{sec:interpretation}.

\section{Method for measuring vorticity}  
\label{sec:method}

Measuring vorticity in interstellar turbulence is not straightforward
because (1) the only velocity component accessible to measurement is its
projection on the line of sight ($v_x$), and (2) measured spatial
variations are limited to those in the plane of the sky ($y,z$).  We thus
have access only to incomplete components of the vorticity
($\omega_{\zeta}=\partial v_x /\partial \zeta$ where $\zeta=y$ or $z$).  By
analyzing the outputs of numerical simulations of midly supersonic
turbulence by~\citet{porter94:ksdtdsf}, \citet{lis96:splcvcvict} have shown
that vorticity extrema may be localized in a map of molecular lines with
high enough spectral resolution.  Instead of computing the vorticity (an
essentially inaccessible quantity in astronomy), they study statistical
properties of line centroid velocity increments. More precisely, the
centroid velocity (\Centroid) of a line is its first order moment
\begin{equation} 
  \Centroid = \int T(v_x) v_x \, dv_x \, / \int T(v_x) \, dv_x.
\end{equation}
It is only in the optically thin case, that a line can be interpreted as
the probability distribution function of the radial velocity along the
observed line of sight. In this case, the centroid velocity is just the
column density weighted average of the velocity along the line of sight.
Centroid velocity increments associated to a lag \slag, are the differences
of centroid velocities for any pair of lines of sight separated by the
distance \slag{} in the plane of sky. Thus, if \r{} is the vector position
and \vlag{} is a vector of modulus \slag{} lying in the plane of sky,
centroid velocity increments are defined as
\begin{equation}
  \label{eq:cvid}
  \Cincr(\r,\vlag) \Definition \Centroid(\r+\vlag) - \Centroid(\r).
\end{equation}
\citet{lis96:splcvcvict} show that the PDFs of centroid velocity increments
have non-Gaussian wings at small lags, those wings gradually disappearing
when the lag increases. More interestingly, they show that the spatial
distribution of the positions populating these non-Gaussian wings is the
same as that of the extrema of $\langle (\nabla \times {\bf v})_y \rangle
^2 + \langle (\nabla \times {\bf v})_z \rangle ^2$, where the brackets hold
for the line--of--sight average. In the optically thin case, this result
simply follows from the linearity of the operations of derivation,
difference and summation along a line of sight.  The largest centroid
increments thus trace a subset of the regions of largest vorticity,
integrated along the line of sight. It has to be appreciated here that the
averaging along the line of sight occurs first, over vorticity components
which have a sign.  An extremum of the above quantity means the occurrence
along the line of sight of one (or a few) events far above the average.

This method has been challenged by~\citet{klessen00:opPDFstfsgm} who
computes the PDFs of line centroid increments ($\Delta v$-PDFs) in
numerical simulations of compressible turbulence with and without
self-gravity.  He finds that, when computed in simulations of decaying
turbulence, $\Delta v$-PDFs have shapes in disagreement with those
observed. He shows that the inclusion of self-gravity in the simulations
leads to better agreement with the $\Delta v$-PDFs observed in molecular
clouds.  This discussion is valuable but its bearing is somewhat limited by
the small size of the simulations (64$^3$). In addition, the confrontation
with the observations is restricted to star forming regions. It is
therefore not surprising that the agreement with the numerical simulations
be better when self-gravity is included in the simulations.  Klessen
concludes that "one should not rely on analyzing velocity PDFs alone to
disentangle the different physical processes influencing interstellar
turbulence". We fully agree with him.  Our scope here is far more limited.
We search for regions of enhanced vorticity in turbulent interstellar
clouds far from star forming regions, with the perspective of tracing
bursts of dissipation of turbulence and characterizing them, on the basis
of their statistical properties, spatial distribution and morphology.

The spectral correlation function (SCF) introduced
by~\citet{rosolowsky99:scf:ntaslm} is a general tool that is able to detect
velocity and/or column density variations through an analysis of line shape
differences. The flexibility of the SCF formalism makes it easy to use for
direct comparison between observations and
simulations~\citep{padoan03:scfmc:sttm}. For instance,
\cite{ballesteros02:vsismsscf} use the local form of the SCF to detect
small scale velocity structures in \HI{} observations of the north
celestial pole loop.  However, as recognized by those authors, this method
cannot disentangle between the various origins of the small scale velocity
variations (\eg{} caustics versus shocks). The method we propose here is
complementary as centroid velocity increments directly relate to the
velocity shear (and thus vorticity) in the optically thin case.

\section{Description of the observed fields}
\label{sec:fields}

The fields analyzed in this paper fulfill several requirements: (1) the
observed maps are large enough to provide significant statistics on the
velocity field, (2) the fields are far from star forming regions, in order
not to include internal energy sources (outflows, HII regions in expansion,
...) in the statistics, (3) they comprise low mass, almost thermally
supported dense cores.

The two fields are nearby regions of low average column density, except in
the small projection area of the dense cores.  One, MCLD123.5+24.9
(hereafter referred to as Polaris), is located in the high latitude cirrus
cloud of the Polaris Flare and the other, (hereafter L1512), is located at
the eastern edge of the Taurus-Auriga complex. The average H$_2$ column
densities (at the arc min scale) deduced from \twCO{} observations and/or
star counts~\citep{cambresy01:firdovepf} in these fields are respectively
$N({\rm H}_2) \sim 10^{21}$ and 2.5 $\times 10^{20}$\pscm{} or about 1 and
0.25 mag of visual extinction, respectively.  For comparison, the visual
extinction in the Taurus molecular cloud reaches 33 mag in the most opaque
regions like TMC1 or L1495~\citep{padoan02:sfsemt}, about one hundred times
more opaque than the transparent areas that we analyze in this paper.

These fields have been mapped in the two lowest rotational transitions of
\twCO{} and \thCO{} as well as in \CeiO{} \Jone{} at the IRAM-30m
telescope~\citep{falgarone98:ikp:ssspsfr}.  A larger field around L1512 has
been mapped with the 10.4 m antenna of the Caltech Submillimeter
Observatory (CSO) in the \twCO{} \Jtwo{} line and is described
in~\citet{falgarone01:fshmfesdc}.  The characteristics of the observations
are summarized in Table~\ref{tab:tels} and~\ref{tab:fields}. Note that all
the maps are Nyquist sampled or better. The average signal-to-noise ratios
(and their rms dispersion within the maps) are given in the last two
columns of Table~\ref{tab:fields}, for the line peak temperatures and the
line integrated areas.  These numbers illustrate the high quality of the
data.  The maps of the line integrated areas (zero order moment), and line
centroid velocities (first order moment) are displayed in
Figs.~\ref{fig:sig:pol} through~\ref{fig:sig:l1512s}.  We have inserted the
high angular resolution IRAM data at their relevant position in the large
scale CSO field of L1512 to illustrate the agreement between the two data
sets (Fig.~\ref{fig:sig:l1512s}).

\TableTelDescription{} %
\TableFieldDescription{} %
\FigureSignalPolaris{} %
\FigureSignalLFifteenTwelveIRAM{} %
\FigureSignalLFifteenTwelveCSO{} %
\FigureDenseCores{} %

In spite of their low average extinction, each field harbours a low mass
dense core.  The dense core in the Polaris field has been mapped in the
CS(2-1), (3-2) and (5-4) lines \citep{heithausen99:eimgcc} and in a number
of molecular species at a few positions only by \citet{gerin97:ciic:hlc}.
Recent observations of the dust thermal
emission~\citep{heithausen02:gbcmcc} show a clear peak of submillimeter
dust continuum emission coinciding with the \CeiO{} peak
of~\citet{falgarone98:ikp:ssspsfr}.  Well-defined and barely resolved peaks
of HC$_3$N emission are found to be clearly displaced relative to the dust
continuum maximum.  They are indicated by two crosses in the \thCO{} map of
Fig.~\ref{fig:sig:pol}. \HH{} densities as large as a few 10$^5$\pccm{} in
this dense core are derived from the multi-line analysis of
\citet{heithausen02:gbcmcc} and \citet{gerin97:ciic:hlc}.  Our unpublished
map of CS(3-2) integrated emission is displayed in Fig.~\ref{fig:dc} and
delineates the region of largest density. The contour level at 0.6 \Kkms of
the CS(3-2) emission is shown in Fig.~\ref{fig:sig:pol} to help localize
the dense core.  The \thCO\Jone{} contour level of 4 \Kkms{} is also drawn
on Fig.~\ref{fig:sig:pol} to indicate the region of column density larger
than $\sim 4 \times 10^{21}$\pscm.
 
In L1512, the dense core has been mapped in CS(2-1) by~\citet{fuller89}
over several arc minutes.  \citet{lee01:imtsc} have mapped it over a more
restricted area in the CS(2-1) and N$_2$H$^+$ lines.  A map of the
\HCOp(3-2) emission has been performed at the CSO in November 2002. The
\HCOp(3-2) line integrated emission is displayed in Fig.~\ref{fig:dc}.  As
in other dense cores, the N$_2$H$^+$ emission is more concentrated than the
CS emission and the N$_2$H$^+$ and \HCOp(3-2) emissions have very similar
boundaries.  Again, in the following, we adopt the 0.15\Kkms{} contour of
the \HCOp(3-2) emission as the boundary of the L1512 dense core.  It is
drawn in Figs.~\ref{fig:sig:l1512} and~\ref{fig:sig:l1512s} to help locate
the dense core.  The maps of Figs.~\ref{fig:sig:pol}
to~\ref{fig:sig:l1512s} confirm that dense cores, as traced by molecular
lines such as CS, N$_2$H$^+$ or \HCOp{} at millimeter wavelengths, are
embedded in larger structures of moderate column density, traced by the
\thCO{} lines, and further down in column density, by the \twCO{} lines.
Note that the \twCO{} integrated emission in both fields is far from being
isotropically distributed around the dense cores.

The analysis presented in the next sections has been conducted on the
\twCO{} and \thCO{} data sets, when available. \twCO{} is the only
available tracer of molecular gas down to \HH\ column densities as low as a
few 10$^{20}$\pscm. Reaching these limits is mandatory in order to be able
to make a link with statistical works performed on HI data ({\it
  e.g.}~\citet{miville-deschenes03,brunt03:utmimede}), and to obtain large
maps avoiding star forming regions. However, the large opacity of the
\twCO{} line prevents the sampling of all the gas on the line of sight,
most severely in regions close to the dense cores. The \thCO{} lines are
thus used to analyze the gas in regions of intermediate column density
between the transparent cloud edges traced by \twCO{} only and the dense
cores.  The areas available for statistical analysis in this line are
smaller than those available in the \twCO{} line because of the different
abundances of the two isotopomers. The two lines are therefore
complementary.

\section{Statistical analysis}
\label{sec:stat}

\subsection{Computation of line centroids and their increments}
\label{sec:stat:computation}

\FigureSpectraMap{} %

Centroid velocities have been computed according to the algorithm described
in~\citet{pety99:a:dsitks}. The line window used to compute the line
centroid is critical (see Appendix~\ref{sec:noise:snr}) and we have
adjusted it locally to maximize the signal--to--noise ratio of the
integrated area defined as \SNR[a]=$\Sigma_1^m T_i/\sqrt{m} \sigma$ where
$T_i$ are the temperatures of the $m$ channels within the window and
$\sigma$ the rms noise level of the spectrum.  As the optimal window is
expected to vary softly from one position to another, we smoothed the
variations of the window edges with a median, moving boxcar filter of size
$5\times 5$ pixels. The first moment is then computed on each spectrum
using the optimal window found for its location.  Fig.~\ref{fig:smp:l1512}
illustrates the optimal windows found by this algorithm for a subset of the
CSO map.  This method also ensures that there is no significant emission
out of the final optimal window.  The line centroid increments are then
computed over a lag $l$ expressed in pixels of the map. These increments
are computed only between spectra that \emph{(i)} have a \SNR[a] greater
than 10.0 and \emph{(ii)} have at least 4 neighbors with $\SNR[a] \ge
10.0$. We show in Appendix~\ref{sec:noise:simul} how this threshold \SNR[a]
= 10.0 ensures that our statistical analysis of velocity increments is not
contaminated by thermal noise.  For a given lag $l$, the increments are
computed for half the possible orientations of the lag vector because if
all the orientations are kept for each lag, each increment in the map
appears twice with opposite signs. In the case of $l=3$ pixels, for
instance, there are 16 different orientations, defined by the 16 neighbors
of any position lying within the rings of radii 2.5 and 3.5 pixels. We kept
only 9 different orientations for each position in the maps.  Note that the
smallest significant lag is $l=2$ pixels for the CSO map and $l=3$ pixels
for the IRAM maps because the maps are oversampled.

\subsection{Non-Gaussian wings in the centroid increment PDFs.}
\label{sec:stat:non-gauss-wings}
 
\TableNormalization{} %
\FigurePDFsPolaris{} %
\FigurePDFsLFifteenTwelve{} %

We have built the PDFs of the computed centroid increments for four
different lags in each field and for each line. Hereafter, the PDFs of
centroid increments will be called \Cincr{}--PDFs or increment PDFs. They
may also be seen as shear--PDFs since, as said in Section 2, the measured
projection of the velocity is orthogonal to the plane in which
displacements are measured (\ie\ the plane of the sky).

Figs.~\ref{fig:PDFs:pol} and~\ref{fig:PDFs:l1512} show the evolution of the
\Cincr{}--PDFs with increasing values of the lags for each map.  The PDFs
are normalized to zero mean and unit dispersion and a normalized Gaussian
PDF with the same dispersion is shown, for comparison, as a dotted line.
The quantities used for this normalisation are given in
Table~\ref{tab:norm} for each map and a lag of 3 pixels.  $\langle
\Cincr[3] \rangle$ is the offset applied to the increments at lag $l=3$
pixels to get a centered PDF and $\sigma_{\Cincr[3]}$ is the standard
deviation of the increments, for the same lag.  The ratios ${\langle
  \Cincr[3] \rangle \over \sigma_{\Cincr[3]}}$ are given in Col. 5: they
are all significantly smaller than unity.  We therefore feel confident that
the normalisation we apply does not affect the estimate of the
\Cincr{}--PDFs computed at small lags.  The error bars drawn simply reflect
the number of elements in each bin of the PDF.  Note that the statistical
significance of the largest increments is best for the CSO field because
the number of \emph{independent} spectra available in this map is about
nine times larger than in the IRAM maps.  This illustrates the importance
of the size of the maps for this kind of statistical study where departures
from Gaussianity are searched for and occur only with probabilities close
to 10$^{-2}$ or below. In Appendix~\ref{sec:sampling}, we compute the
largest statistically significant lags for each map and two values of the
numbers of bins in the PDFs, 30 and 8 bins. These numbers confirm that PDFs
built with 30 bins, as in Figs.~\ref{fig:PDFs:pol}
and~\ref{fig:PDFs:l1512}, are statistically significant up to lags of
$\sim$ 6 pixels for the small fields and 12 pixels for the large one. They
also show that the broad features in the PDFs, such as non-Gaussian wings,
spreading over 4 bins or more, are significant up to lags of 12 pixels for
all the fields.

All the sets of increment--PDFs exhibit non-Gaussian wings with departure
from Gaussianity being more pronounced as the lag becomes small. The effect
is the most visible in the large scale L1512 field, because the large
number of independent spectra in that field reduces effects due to the poor
sampling at large lags and makes the PDFs there the most symmetrical.

\subsection{The influence of large--scale velocity gradients}
\label{sec:stat:lvg}

\TableLargeVelocityGradient{} %
\FigureSDEVvsLag{} %

Large scale variations of the line centroid velocities are visible on Figs.
1 to 3.  They most likely trace large scale velocity gradients. The
observed values are derived from the two extreme values of the centroid
velocity $\Centroid[min]$ and $\Centroid[max]$ and the size scale over
which they are measured $s$ (Table~\ref{tab:lvg}).  These numbers provide
an estimate of the observed velocity gradients $\ObsGrad \approx
\abs{\Centroid[max]-\Centroid[min]}/s $ given in column 5 of
Table~\ref{tab:lvg}.  We note here that the values derived from two
different lines in the same field are not necessarily the same because the
lines do not quite sample the same gas, as discussed above.  Unlike in
other studies~\citep{grosdidier01:hsti:cfhfps:en:ts,miesch99:vfssfr:cvo},
we did not remove the contribution of large scale velocity gradients in our
centroid increment computations.  In this section, we explain why.
 
In the analysis of turbulence in HII regions, the large scale velocity
gradients are related to the expansion of the warm ionized gas, driven by
the pressure gradient of the HII region relative to the surrounding medium.
It is thus justified to remove large scale gradients, prior to any
statistical analysis of the turbulence within HII regions, as done
by~\citet{grosdidier01:hsti:cfhfps:en:ts}, because the HII region expansion
is an ordered large scale motion and is not part of the inertial range of
the turbulent cascade.

\citet{miesch99:vfssfr:cvo}, who analyze the turbulence in molecular
clouds, have different goals from ours.  Their goal is to extract an
homogeneous and isotropic subset of interstellar turbulence and compare it
to laboratory experiments or direct numerical simulations.  Instead, we
search for regions of enhanced shear {\it at small scales} to localize
bursts of dissipation. We show below that we are not limited in our
analysis by large scale anisotropy.  We first argue, following the results
of numerical simulations of \citet{burkert00:tmcc:tp} that turbulence
itself with its steep power spectrum (\ie{} most of the power is in the
large scales) generates velocity gradients which may be interpreted as
rotation at any scale. We show in Fig.~\ref{fig:sdev-vs-lag} and
Table~\ref{tab:lvg} that this is indeed the case.  The scaling of the
standard deviation of the line centroid increments $\sigma_{\Cincr{}_l}$
with lag $l$ is shown in Fig.~\ref{fig:sdev-vs-lag}. Up to lags of $\sim
100$ arc sec or about 12 pixels, $\sigma_{\Cincr{}_l} \propto l^{\zeta_2}$.
The values of $\zeta_2$ are given in Table~\ref{tab:lvg}.  This scaling
provides that of the turbulent velocity shear $\TurbGrad \propto
l^{\zeta_2-1}$ with $l$.  The turbulent velocity shear, \TurbGrad{},
inferred from this scaling law at the scale $s$ of the largest centroid
differences is given in col. 9 of Table~\ref{tab:lvg}.  Comparison of
columns 5 and 9 shows that in all cases: $\ObsGrad \la \TurbGrad$.  The
large scale observed velocity gradients are therefore smaller than or
comparable to those expected from the turbulent shear at the same scale,
suggesting that they are part of the turbulent dynamics.
 
Last, we show that the large--scale gradients do not affect the small scale
statistics which are of importance to the present analysis.  The
contribution of the large--scale gradients to the centroid velocity
increments computed as $\delta v_3 = \ObsGrad \times l$ for $l=3$ pixels
are given in Table~\ref{tab:lvg} for each map (col. 6).  These values are
then compared to the standard deviation of the \Cincr{}--PDFs distribution
(col. 7).  In all cases, the ratio $\delta v_3 / \sigma_{\Cincr[3]}$ is
smaller than or equal to unity. The contribution of the large--scale
gradients to the centroid increments at small scale is therefore much
smaller than the increments populating the non-Gaussian wings which extend
up to 4 or 5 $\sigma_{\Cincr}$ (see Figs. 6 and 7).  Furthermore, the large
scale gradient has a well-defined direction and therefore preferentially
affects the increments computed along the same direction \ie{} one
increment out of 9 in the case of $l=3$ pixels (out of more for larger
lags). The contribution of the large scale velocity gradient to increments
computed over lags of different orientations is further reduced by the
appropriate cosine.  For these reasons, we are confident that keeping the
large scale velocity gradients in the maps does not significantly distort
the \Cincr{}--PDFs of Figs.~\ref{fig:PDFs:pol} and~\ref{fig:PDFs:l1512}.

In summary, we did not remove large--scale velocity gradients prior to our
statistical analysis for two reasons: (1) there is evidence for the
large--scale gradients to be part of the turbulent dynamics we analyze and
(2) their value is such that their contamination of the small scale
statistics on which we focus here, is negligible.

\section{Spatial distribution of the non-Gaussian centroid increments:
  a new kind of small scale structure}
\label{sec:spatial}

\subsection{The positions populating the non-Gaussian wings of the
  increments PDFs are not randomly distributed: they form elongated
  structures}
\label{sec:spatial:not-random}

\TableCVIsStatistics{} %
\FigureCVIMapPolaris{} %
\FigureCVIMapLFifteenTwelveIRAM{} %
\FigureCVIMapLFifteenTwelveCSO{} %
\FigureZoomLFifteenTwelve{} %

We have seen in the previous section that, in most cases, the departure of
the increment PDFs from a Gaussian distribution is as large as the lag is
small.  We are here interested in tracing the locations of the positions
for which the small scale increment values build up the non-Gaussian wings.
We display in Figs.~\ref{fig:cvi:pol} through~\ref{fig:cvi:l1512s} the
spatial distribution of the increments computed for lags of three pixels.
The lag value $l=3$ pixels is the smallest significant lag in the IRAM maps
because the maps are Nyquist sampled or better: adjacent pixels are
therefore not independent (see Table~\ref{tab:tels}).

The spatial distributions of the centroid increments are shown as azimuthal
averages over the 16 possible directions because it allows a more compact
presentation of the results. In the rest of the paper, we note \AveCincr{}
the azimuthal average of the modulus of all the oriented centroid
increments computed at a given position, for lags of three pixels.  The
thresholds, \AveCincr[0], given in Table~\ref{tab:cvi-stat}, has been
chosen to isolate the regions where more than half of the oriented
increments belong to the non-Gaussian wings of the \Cincr{}--PDFs.  These
thresholds are close to the standard deviations $\sigma_{\Cincr[3]}$ of the
increment--PDFs for $l=3$ pixels.  Note that the averaging produces
increment values smaller by factors up to a few than the original values
used to build the PDFs. This effect is illustrated in the cuts of
Figs.~\ref{fig:cut:pol:dec}--\ref{fig:cut:l1512s:ra} where the variations
of the centroid and of the centroid increments before and after azimuthal
averaging are shown as a function of position.

In Figs.~\ref{fig:cvi:pol}--\ref{fig:cvi:l1512s}, the uniform black areas
correspond to positions where increments are not computed because at least
one of two spectra used either has a \SNR[a] value below 10.0 or is too
isolated (\ie{} with less that 4 contiguous neighbors with $\SNR[a] >
10.0$).  The black contours drawn after the CS(3-2) or \HCOp(3-2) maps help
locate the dense cores.

In the sketches of Figs.~\ref{fig:cvi:pol}--\ref{fig:cvi:l1512s}, the black
contours labelled by letters follow the thresholds \AveCincr[0].  These
contours are therefore intended to trace the structures delineated by the
positions where the original, non averaged, increments are larger than a
few $\sigma_{\Cincr[3]}$ and belong to the non-Gaussian wings of the
\Cincr{}--PDFs.

Two characteristics of these structures may be derived from
Figs.~\ref{fig:cvi:pol}--\ref{fig:cvi:l1512s}.  Firstly, the spatial
distribution of the largest averaged increments is not random.  They are
concentrated in specific regions and appear to form elongated structures,
almost straight in several cases, and covering only a small fraction of the
maps.  Secondly, the \AveCincr{} values in these structures are well above
the background fluctuations of the increments in the maps.
Table~\ref{tab:cvi-stat} gives the background values \AveCincr[bg] computed
as the median of all the values smaller than the thresholds \AveCincr[0]
and the largest values of increments \AveCincr[max] in the maps.  It shows
that the largest values are between 4 and 9 times above the background
values, far above statistical fluctuations.

Last, it is remarkable that similar spatial distributions are found for the
large increments computed in the same field with two independent data sets
\ie\ from two different telescopes and two different lines, \twCO{} \Jone{}
and \Jtwo{}. Fig.~\ref{fig:zoom:l1512} allows a comparison of the patterns
found for the centroid increments in the L1512 field from the IRAM-30m and
CSO maps.  The morphological and quantitative agreement between the two
maps fully supports the reality of these structures. In Section 6, we
discuss why they do trace the shear of the velocity field.

The smallest lags of three pixels introduce an effective resolution,
ascribing a size of three pixels to unresolved variations of the line
centroid.  But some of the structures traced by the largest increments are
resolved by the observations.  This last characteristic is also seen in the
cuts of Figs.~\ref{fig:cut:pol:dec}--\ref{fig:cut:l1512s:ra}.  In the RA
cut across the L1512 field, the peak of increment visible at RA offset
$\sim 200"$ is due to the steepening of the variation of the line centroid
over 6 consecutive pixels, from offset $\sim 300"$ to 200" or more than
three CSO beamwidths. The half-power width of this feature (above the
background value), $\sim 0.05$ pc, is therefore resolved by the procedure
of computing increments over lags of 3 pixels. We illustrate with the other
cuts that several of these structures are actually resolved, with
thicknesses of the order of 0.05 pc, up to 0.08 pc.

\FigureCVICutPolarisDec{} %
\FigureCVICutPolarisRA{} %
\FigureCVICutLFifteenTwelveCSODec{} %
\FigureCVICutLFifteenTwelveCSORA{} %
\clearpage{}

\subsection{Comparison with the results of~\citet{miesch99:vfssfr:cvo}}
\label{sec:spatial:comp-MSB}

The non-random distribution of the largest centroid velocity increments
seems at odd with the spotty distribution found
by~\citet{miesch99:vfssfr:cvo} in their maps of centroid velocity
differences of several star forming regions.  However, we consider that our
results agree with theirs for the following reason.  Their method, as
explained in Section~\ref{sec:stat:lvg}, is different from ours because
they first remove a smoothed value from the line centroid values in their
maps. The centroid velocity fluctuations they obtain with this procedure
are already of the same nature as a velocity increment, as shown
by~\citet{lis96:splcvcvict}. It is therefore meaningful to compare their
maps of centroid fluctuations with our maps of centroid increments. Their
maps of centroid velocity fluctuations (their Figs.~4 and~5) do indeed
exhibit spatial structures. The largest values appear in several cases to
form narrow and elongated structures similar to those we find.  The spotty
distribution is found only once they compute the increments of these former
maps of velocity fluctuations. What they call centroid velocity differences
cannot be compared to what we compute.

A more detailed comparison of the two sets of results would not be
meaningful for several reasons.  The signal-to-noise ratios of our data
given in Table~\ref{tab:fields} are significantly better than those
of~\citet{miesch99:vfssfr:cvo}. Their SNR values range between 4.5 and 8.9,
except for HH83, which has SNR=19.7 but is their smallest field.  The
nature of the fields is also different. While we analyze only quiescent
fields in the present paper, Miesch et al. have analyzed active star
forming regions where many shocks, driven by the young stars, interact and
probably generate considerable small scale structure in the patterns of the
vorticity. Last, most of the fields being at distances larger that the
nearby fields studied here, the angular characteristic scales of the
structures (if any) are expected to be smaller.

\subsection{The regions of largest centroid increments are not
  coinciding with density nor column density peaks}
\label{sec:spatial:pure-vel}

\FigureWCOvsCVI{} %

The sketches of Figs.~\ref{fig:cvi:pol} to~\ref{fig:cvi:l1512s} allow a
comparison of the structures of largest increments with the concentrations
of matter, whether they are large column density or large density
structures or both.  The dotted contours delineate the dense cores and the
grey areas those where the \thCO\Jone{} integrated emission is larger than
3 and 4 \Kkms{}, for L1512 and Polaris respectively, which corresponds to
$N(\HH) > $ a few $10^{21}$\pscm, about 10 times the average column density
of the large scale environment.  The main result visible in those figures
is that none of the regions of largest increments coincides, even in
projection, with either a dense core (\ie\ a density peak) or a column
density peak.

This result is better seen in the cuts of
Figs.~\ref{fig:cut:pol:dec}--\ref{fig:cut:l1512s:ra}: the large variations
at small scale of the line centroid, at the origin of the large increment
values, do not coincide with column density extrema.  Such a coincidence
would be expected if the line centroid variations were due to the overlap
on the line of sight of unrelated gas components. In that case, an extremum
of column density would coincide with the region of large centroid
increment, which is that over which the two components are intercepted,
producing a modified (locally broader) line profile.  This lack of
correlation is also seen in the scatter plots of Fig.~\ref{fig:wco-vs-cvi}
where the largest increments are in no case associated with extrema of
column density.  Conversely, the largest increments tend to occur at the
positions were the integrated areas are the smallest.

\subsection{Link between the orientation of these structures and that
  of density structures}

The previous section shows that the structures traced by the regions of
largest centroid increments at small scale do not coincide (in projection)
with the dense cores nor the column density extrema in the maps.

Nonetheless, this statement does not mean that these small scale structures
are unrelated to the dense cores.  The orientation of the most elongated
structures tend to be related with that of the dense cores. In the Polaris
field, the structure labelled C' in the \twCO{} map, appears as a
prolongation of the bright \thCO{} structure, which seems to be an
extension of the dense core itself (see sketches of
Fig.~\ref{fig:cvi:pol}).  The structures B and D' are adjacent and parallel
to the edge of the dense core, as is seen also in the cuts of
Fig.~\ref{fig:cut:pol:ra}.  In the L1512 field (Fig.~\ref{fig:cvi:l1512s}),
similar patterns are visible. The two southern structures A and A' frame in
projection the region of bright \twCO{} lines (dot-dashed contour), and the
structure B is elongated in the same direction as that of the filament of
matter seen in \thCO{} and \twCO\ and the orientation of the dense core
itself as mapped in \HCOp{}.  Since the velocity is continuous between the
structures of large increments and those of matter (see maps of line
centroids in Figs.~\ref{fig:sig:pol}--\ref{fig:sig:l1512s}), it is most
likely that they are connected in the 3--dimensional space.

\subsection{Volume filling factor of the largest increment structures}
\label{sec:spatial:contrast}

For each field, in addition to the maximum value of the increments in the
map, \AveCincr[max] and the background value \AveCincr[bg] defined in
Section~\ref{sec:spatial:not-random}, Table~\ref{tab:cvi-stat} also gathers
the fraction $f_s$ of pixels in the map where the centroid increments are
larger than the thresholds \AveCincr[0].  $f_s$ may be understood as their
surface filling factor in the maps. From its values, one derives volume
filling factors of the regions of largest increments of at most a few \%.
Assuming that the depth of the structures along the line of sight is
comparable to their projected thickness, $d$, and that the extent of the
cloud along the line of sight, $L$, is provided by the large scale
maps~\citep{falgarone98:ikp:ssspsfr}, $f_v \about f_s d/L$. For $d\sim
0.05$ pc and $L\sim 1$ pc, $f_v \sim$ a few \% or less.

\section{Discussion}
\label{sec:discussion}

The subset of spectra shown in Fig.~\ref{fig:smp:l1512} encompasses one of
the regions where the centroid increments are the largest.
Fig.~\ref{fig:smp:l1512} shows that the optimal window found to compute the
line centroids varies only by very small amounts from one spectrum to the
next. This illustrates that the measured centroid increments are due to
small variations in the shapes of the line profiles from one line of sight
to the next. These variations may have several origins, unrelated to the
structure of the velocity field: optical depth effects, gas temperature or
density fluctuations. We show in the following section that our
interpretation of these variations as tracers of small scale velocity
shears is justified.

\subsection{Optical depth effects}
\label{sec:disc:opt}

The optical depth of the lines used introduces a possible bias in the
sampling of the velocity field performed by the spectra. Indeed, optically
thick lines tend to sample preferentially foreground material.

However, the maps of centroid increments built from the \thCO{} and \twCO{}
lines, although not exactly the same, are reminiscent of each other,
suggesting that the two lines sample about the same gas (see
Figs~\ref{fig:cvi:pol}--\ref{fig:cvi:l1512}).  This is also suggested by
the similarity of the maps obtained in L1512 with two different \twCO{}
transitions of different optical depths (Fig.~\ref{fig:zoom:l1512}). These
results support the concept of low effective optical depth of the \twCO\ 
lines discussed in~\citet{martin84:coefmc}.
\citet{falgarone98:ikp:ssspsfr} also inferred a low effective optical depth
in the \twCO{} lines of the fields analyzed here (except in the projected
area of the dense cores), on the basis of the uniformity of the excitation
temperature. Note that a low effective optical depth of the \twCO{} lines
is to be expected in these fields (as opposed to brighter molecular clouds)
because a large velocity dispersion of the gas, leading to a large escape
probability of the photons in velocity space, is combined with low gas
column densities.  The same similarity is found between maps of centroid
increments built from the \thCO{} and \CeiO{} spectra of the environment of
the L1689B dense core (Pety et al. in preparation), suggesting that optical
depth effects are not seriously affecting the centroid increment
determinations.

This result illustrates a major strength of the line centroid analysis. On
the one hand, centroids are velocity moments and as such provide more
weight to velocities far from the line centroids, i.e. the line wings.  The
wings are the parts of the line profiles the least affected by optical
depth effects.  On the other hand, our method which limits the range of
velocities used to compute the centroid (see
Section~\ref{sec:stat:computation}), limits the noise contribution,
particularly in the line wings.  We thus argue that there is a velocity
range in spectrally well-sampled line profiles, neither too far in the line
wings (to avoid noise artefacts) nor too close to the line centroid (to
avoid optical depth artefacts), which carries enough information on the
velocity field to be used on statistical grounds.

Last, self-absorption is also unlikely in those low brightness core
environments because the excitation temperature of the \twCO{} and \thCO{}
lines is remarkably uniform as was shown by the scatter plots of the
\Jone{} to \Jtwo{} \twCO{} and \thCO{}
lines~\citep{falgarone98:ikp:ssspsfr}. For self-absorption to occur the
optical depth must exceed unity and the foreground material must have a
lower excitation temperature.

\subsection{Role of temperature fluctuations}
\label{sec:disc:temp}

It might be argued that the variations of line centroids traced by the
\twCO{} lines are due only to variations of the gas kinetic temperature.
This might be the case in bright sources where the \twCO{} lines are
thermalized. But it has to be appreciated that in the fields under study
here the \twCO\ lines are weak (less than 5 K) while the gas is poorly
shielded from the ambient UV field (\HH\ column densities less than a few
$10^{21}$\pscm{}) and thus likely to be much warmer than 5K.  It means that
either the lines are subthermally excited, in which case their excitation
is mostly radiative, or that they are thermally excited in small structures
with a large beam dilution factor. In either case, the line profile
variations are not determined by variations of the gas kinetic temperature
but by those of the radiative excitation and/or beam filling factor, which
both depend on the velocity field~\citep{falgarone98:ikp:ssspsfr}.

\subsection{Role of density fluctuations unrelated to the velocity field}
\label{sec:disc:dens}

\FigureRAIvsCVI{} %

In this section, we discuss the influence of column density or density
variations, independent of the velocity field, on the velocity increments
that we measure.  Under the optically thin hypothesis, $dN(u) \propto
T(u)\, du$ where $dN(u)$ is the column density of gas at the projected
velocity $u$ within $du$.  The centroid velocity defined in Eq. (1) is
therefore also

\begin{equation}
  \Centroid = \frac{ \int u \,dN(u)}{\Ntot}
  \quad \mbox{with} \quad
  \Ntot = \int dN(u).
\end{equation}
The centroid increment between lines of sight 1 and 2 can thus be written
\begin{equation}
  \label{eq:3}
  \Cincr[1-2] =   \bracket{\frac{\int u_1 dN_1(u_1) }{\Ntot_1}}
                   - \bracket{\frac{\int u_2 dN_2(u_2)}{\Ntot_2}}.
\end{equation}

From this expression, it is obvious that $\Cincr[1-2] = 0$ if the two lines
of sight differ only by their total column density, the velocity structure
being the same \ie{} $ N_2(u_2) \propto N_1(u_1)$ for all velocities. In
this case the two integrated line profiles are simply homothetic.

Let us consider another simple case in which the total column density $N$
along the two lines of sight is approximately the same, the only difference
being that a fraction of the gas at projected velocity $u_1$ on the first
line of sight is at $u_2$ on the other. This is possible, for instance, in
the perspective of matter distributed in a large number of very small
structures, possibly virialized in the potential well of the cloud.  Then
$\Cincr[1-2]= (u_1-u_2)\,\delta N/N$ where $\delta N$ is the column density
of the gas which is at different projected velocities on each line of
sight.  We estimate below the value of $\delta N/N$ required to produce
centroid increments of the order of \AveCincr[max] in each field.  We
assume that the velocity difference $u_1-u_2$ is of the order of magnitude
of the average linewidth in each field $\langle \Delta v \rangle $ given in
Table~\ref{tab:cvi-stat}, $u_1-u_2 \sim \langle \Delta v \rangle $.
Therefore an estimate of the required column density fluctuations is:
$\delta N/N \sim \AveCincr[max]/\langle \Delta v \rangle$.  The latter
ratio can be computed from the values in Table~\ref{tab:cvi-stat} and is
equal to 1.25, 0.5 and 1.05 for the three fields of
Fig.~\ref{fig:rai-vs-cvi}, respectively.  Now, if we assume that the
relative variations of the line integrated areas trace those of gas column
density, we can check how the observed fluctuations of column density
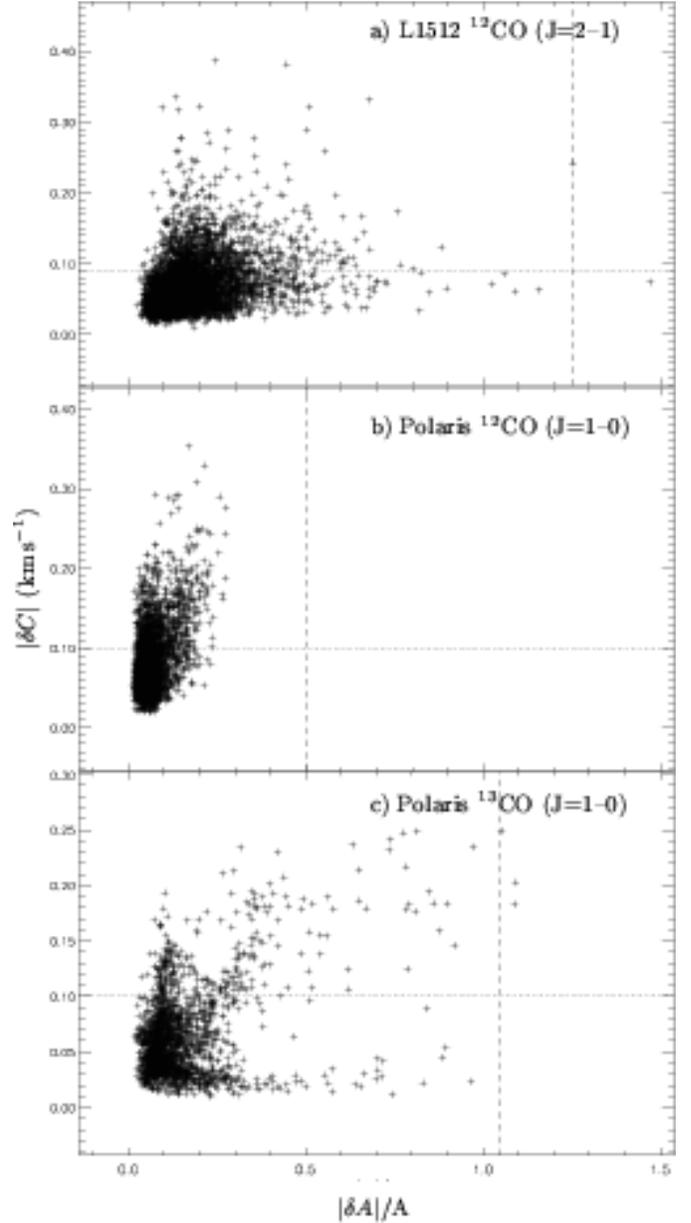
compare with the above values. This may be seen in the scatter plots of the
averaged increments versus the relative variations of the line integrated
area, $\abs{\delta A}/A$ (Fig.~\ref{fig:rai-vs-cvi}).  These scatter plots
are shown only for the three maps where the statistics are the least
affected by the dense core.  The values of $\abs{\delta
  \Centroid[max]}/\langle \Delta v \rangle$ and of the threshold
$\abs{\delta C_0}$ are indicated for each field.  The scatter plots show
that, except for a few lines of sight of the Polaris \thCO\ map, the
observed relative variations of column density (or $\abs{\delta A}/A$) at
the positions of increments populating the non-Gaussian wings ({\it i.e.}
increments above the thresholds \AveCincr[0]) are smaller than the minimum
value estimated above. In the large scale L1512 field and the Polaris field
as seen in \twCO, it is therefore unlikely that the observed centroid
increments trace small scale column density fluctuations due for instance
to randomly distributed clumps.  This is also supported by the fact that
the structures traced by the large centroid increments form spatially
coherent patterns and are not randomly distributed as would be expected for
randomly moving clumps.  We thus conclude that the contribution of small
scale random column density variations to the observed centroid increments
is not dominant.

These small scale column density fluctuations could trace density
fluctuations. They occur over lags of 3 pixels or $l \sim 5000$ AU (see
Table~\ref{tab:tels}). The corresponding density enhancements are therefore
expected to occur over the same small scale $l$. Thus, $\delta n / n \sim
(\abs{\delta A} /A) \, (L/l)$ where $L$ is the depth of the cloud.  For
$\abs{\delta A} /A = 0.5$ and $L = 1 $ pc, $\delta n / n = 25$. These might
be supersonic shocks with Mach numbers of about 5, in which case the
density and velocity fields are closely correlated.  However, such large
density enhancements would produce variations of the CO\Jtwo{}/CO\Jone{}
line ratio in the weak regions of the fields, variations which are not
observed~\citep{falgarone98:ikp:ssspsfr}.

The tenuous variations of the line profiles at the origin of the large
centroid increments at small scale are therefore most likely ascribed to
large velocity shears at small scale.

\section{Possible link with the dissipation of turbulence}
\label{sec:interpretation}

Our statistical analysis of the environment of two low mass dense cores
leads to two main results: {\it (i)} the PDFs of line centroid increments
(or velocity shears) have non-Gaussian wings as more pronounced as the lag
over which they are computed is small, and {\it (ii)} the spatial
distribution of the largest shears is not random and reveals often resolved
and elongated structures not coinciding with density nor column density
peaks. These are predominantly velocity structures.

The first property is a characteristic shared by incompressible turbulence,
compressible turbulence and compressible, magnetized turbulence and is a
signature of the intermittency of the velocity or current density fields.
Our results therefore suggest that the observed distributions of large
shears may trace regions of intense vorticity reminiscent of those
responsible for the intermittency of turbulence.  Note that these regions
of intense vorticity are embedded in the bulk of the flow and that the
large centroid increments are only tracers of their presence in the flow.
In other words, a small scale structure of large vorticity, embedded in the
bulk of the flow, can be at the origin of a small variation of the line
centroids, along the boundary of two large scale regions of more uniform
centroid velocity, \ie{} the large eddies.

We have argued that the large shears are unlikely to trace shocks but we
cannot rule out the fact that the vorticity extrema are fossil structures
of shocks, already disrupted.

If the largest shears that we have detected trace local enhancements of the
vorticity and therefore dissipation, as discussed in the Introduction, an
estimate of the local enhancement of the dissipation rate in those regions
is provided by the ratio $(\Cincr[max]/\Cincr[bg])^2$
(Table~\ref{tab:cvi-stat}) because the dissipation rate follows the square
of the vorticity (or of the velocity shear) (Landau \& Lifshitz, 1987).
The local enhancements are large, between 15 and 80
(Table~\ref{tab:cvi-stat}).  It is interesting to note that the smallest
values of $(\Cincr[max]/\Cincr[bg])^2$ are obtained in the close vicinity
of the L1512 dense core (IRAM field) and the largest in Polaris and in the
large scale environment of L1512 (CSO field).  This may reflect the fact
that turbulent dissipation is a less violent process within a dense core
which is already formed.  We also note that even the weakest structures in
the maps of \AveCincr{} have a contrast larger than 2 above the background
and therefore correspond to a dissipation rate locally 4 times larger than
in the bulk of the field (see for instance the weak structures, labelled B
and C, in the North of the IRAM L1512 \twCO{} map).

Last, the quantity $f_{\epsilon}= \sum_{\AveCincr{} > \AveCincr[0]}
\AveCincr{}^2/\sum_\emr{all}\AveCincr{}^2$ (Table~\ref{tab:cvi-stat}) may
be understood as the fraction of the dissipation taking place in the
regions of large increments (those populating the non-Gaussian wings).  The
underlying assumption is that the ratio of the actual dissipation rate to
$\AveCincr{}^2$ is the same in the regions where the increments populate
the Gaussian cores of the PDFs (the bulk of the fields) as in those
populating the non-Gaussian wings.  Under this assumption, the values of
Table~\ref{tab:cvi-stat} show that a significant fraction (more than half)
of the total dissipation takes place in the ensemble of small scale
structures filling a volume close to a few \% of the cloud volume, as
estimated above.

\section{Conclusion}

We have found a new kind of small scale structure in the environment of low
mass starless dense cores.

They emerge in the maps of line centroid increments between adjacent
spectra as regions between 4 and 9 times brighter than the background
fluctuations.  They are essentially velocity structures because they are
not coinciding with any detected sufficient increase in the column density
or density.  They are small scale, often elongated, structures. A few are
resolved by the observations with thicknesses up to 0.08 pc. Most of them
are not resolved \ie\ smaller than 0.02 pc, the effective resolution due to
the smallest lag available to compute the increments.

The measured projected velocity being orthogonal to the plane of the sky,
in which lags are measured, the centroid velocity increments trace one
component of the shear of the velocity field.  The shear-PDFs exhibit
non-Gaussian wings which are the most prominent for the smallest lag. For
this reason, we ascribe these structures to turbulence by analogy with the
behavior of such PDFs in laboratory flows or numerical simulations of
compressible turbulence.  The new structures delineate the locus of
positions where the increments are the largest and populate the
non-Gaussian wings.

These structures therefore possibly trace regions of enhanced dissipation
of turbulence in the observed fields. The nature of the dissipation process
is not determined by the observations: it could be either viscous
dissipation, ion-neutral collisions or enhanced current densities.  These
structures are unlikely to trace shocks, but they may trace the remanent
vorticity generated by shocks already disrupted. In this picture, a
significant fraction of the turbulent energy present in the field would be
dissipating in those structures filling less than a few \% of the cloud
volume.

Last, the structures display some connexion with the tracers of dense gas,
such as shared orientations and velocity and space pattern continuity.
Before we can make a link between such structures and the formation of
dense cores, we need to investigate control fields \ie{} molecular clouds
without dense cores (Hily-Blant et al. in preparation) and with moderate
star formation activity (Pety et al., in preparation).

\acknowledgements{We thank our referee, Alyssa Goodman, for her stimulating
  comments that helped us significantly improve the paper. We are also
  grateful to Malcolm Walmsley for his careful reading of the manuscript
  and thorough comments.}

\bibliography{ms2555}%
\bibliographystyle{apj}%

\appendix{}

\section{Role of the thermal noise in the statistical analysis}
\label{sec:noise}

Thermal noise in the spectra affects the estimation of the line centroid
velocity. It thus could be that thermal noise largely contributes to the
non-Gaussian wings of the \Cincr{}--PDFs. This influence of thermal noise
is relatively more important at small lags where the increments are the
smallest because of the scaling laws of turbulence. In
section~\ref{sec:noise:snr}, we show that most of the large increment
values are significantly above the thermal noise contribution.  And in
section~\ref{sec:noise:simul}, we present results of numerical simulations
performed to answer two questions: \emph{(i)} can thermal noise create
non-Gaussian wings on intrinsically (originally) Gaussian \Cincr{}--PDFs
and \emph{(ii)} how does thermal noise affect the shape of an \Cincr{}--PDF
which has intrinsic non-Gaussian wings?

\subsection{Increment SNR}
\label{sec:noise:snr}

The uncertainty on the centroid velocity of a spectrum scales
as~\citep{pety99:a:dsitks}:
\begin{equation}
  \Ucentroid \propto \WindowWidth / \SNR[a]
\end{equation}
where $\WindowWidth$ is the width of the optimal window (see
Section~\ref{sec:stat:computation}). The uncertainty on the centroid
velocity increments, $\Delta(\Cincr{})$, can be approximated as the
quadratic sum of the uncertainties on the centroid of each spectrum in the
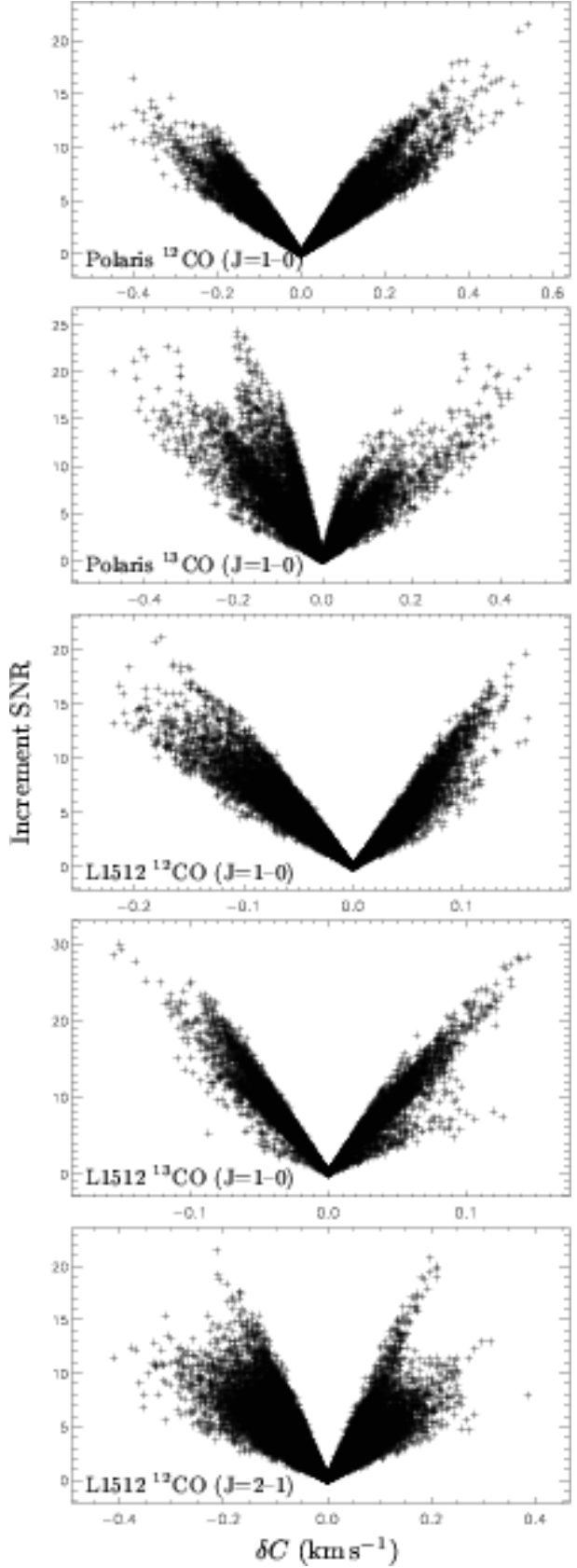
pair.  Fig.~\ref{fig:snr-vs-cvi} shows the scatter plots of the centroid
velocity increments signal--to--noise ratios (\SNR[\Cincr{}], defined as
the ratio of \Cincr{} to its uncertainty) versus the centroid velocity
increments for the five maps. Here all the directions of the pairs are
considered.

The butterfly shape of these scatter plots is due to the fact that the
centroid uncertainties are bounded, essentially by the two extreme values
of \SNR[a], since \WindowWidth{} does not vary much in a map.  We impose
the lowest \SNR[a] value to be 10.0 (see Appendix~\ref{sec:noise:simul}).
The upper value reflects the finite noise level and the peak line intensity
in each map.  The \Cincr{}--uncertainties, $\Delta(\Cincr{})$, defined as
above, are therefore also bounded. We call these boundaries $A$ and $B$ so
that $1/A \leq \Delta(\Cincr) \leq 1/B$. We then obtain
\begin{equation}
  \label{eq:2}
       B*\AveCincr{}
  \leq \SNR[\Cincr{}] \Definition \abs{\frac{\Cincr{}}{\Delta(\Cincr{})}}
  \leq A*\AveCincr{}.
\end{equation}
The $A$ and $B$ boundaries thus appear as the slopes of the lines that
limit the scatter plots of Fig.~\ref{fig:snr-vs-cvi}.

These scatter plots also show that the structures delineated by the largest
centroid increments and shown in Figs.~\ref{fig:cvi:pol}
to~\ref{fig:cvi:l1512s} have large SNRs. Clearly these structures are far
above the noise level, as was already suggested by the good agreement
between the IRAM and CSO results for the L1512 core (see
Fig.~\ref{fig:zoom:l1512} and section~\ref{sec:spatial:not-random}).

\FigureSNRvsCVI{} %

\subsection{Simulations}
\label{sec:noise:simul}

\FigureNormalizedPDFsShapeVSNoise{} %
\FigureNumberOfSpectraVSNoise{} %

We computed two spectra maps made of Gaussian lines with the same line
areas and widths at each position as in the L1512 large scale field. The
first one also has the same spatial distribution of centroid velocities as
the L1512 field. For the second one, the spatial distribution of centroid
velocities has been computed from a random, Gaussian data cube whose power
spectrum is a power law of -5/3 exponent.  It has been checked that this
cube does not show intermittent behavior, \ie{} \Cincr{}--PDF are Gaussian
at all scales~\citep{pety99:a:dsitks}.  These two spectra maps are
respectively referred to as the L1512 velocity field and the Gaussian
velocity field.

To mimic thermal noise, we added random numbers sampled from a Gaussian law
of zero mean and a given standard deviation, to the intensities of all
channels. We checked that the average noise of the spectra in the map is
the value of this standard deviation. The original L1512 CSO field has
quite a uniform noise distribution of average value 0.2\K{}. We thus fixed
the standard deviation of Gaussian noise to values between 0.02 and 0.8\K{}
to explore a fair range of \SNR[a] (the spectra area being kept fixed). We
then used our algorithm to compute the centroid map from the spectra map
(see section~\ref{sec:stat:computation}). We used two different \SNR[a]
thresholds (i.e. 3 and 10) to thoroughly explore thermal noise effects. All
the \Cincr{}--PDFs have been computed for a lag of 3 pixels, and normalized
to zero mean and unit dispersion.

Fig.~\ref{fig:PDFs:norm} shows normalized \Cincr{}--PDFs for 3 different
values of noise levels: 0.002\K{} equivalent to the noise--free case,
0.2\K{} equal to the value of the real L1512 data and 0.8\K{}, the largest
value. For the Gaussian velocity field, thermal noise can create
non-Gaussian wings to the \Cincr{}--PDFs. However, those non-Gaussian wings
appear only when the \SNR[a] threshold is 3 and for the largest noise
values. This implies that thermal noise cannot be responsible for the
non-Gaussian wings seen in the \Cincr{}--PDFs computed on the real L1512
data: the L1512 velocity field has intrinsic non-Gaussian statistical
properties.

For the L1512 Velocity Field (see Fig.~\ref{fig:PDFs:norm}), thermal noise
does not change the normalized \Cincr{}--PDF shape for typical noise values
(\ie{} 0.2\K{}). \Cincr{}--PDF shapes change only for the highest noise
values. In particular, the \Cincr{}--PDF can become Gaussian when the
\SNR[a] threshold is 10.  This is just a sampling effect. Indeed,
Fig.~\ref{fig:npts-vs-noise} shows the number of points in the spectra map
whose \SNR[a] values are above the threshold, as a function of the map
noise level. It can be seen that the number of spectra with $\SNR[a] \ge
10.0$ (solid line) decreases much faster than the number of spectra with
$\SNR[a] \ge 3.0$ (dotted line). Due to the anticorrelation between high
area and high \AveCincr{} regions (see Fig.~\ref{fig:wco-vs-cvi}.a), the
remaining points naturally have a velocity field with a Gaussian statistic.

\section{Sampling effects}
\label{sec:sampling}

\Tablelag{} %

Statistical estimators are significant only if they are measured over a
large enough number of realizations.  The number of realizations of the
increments at a given lag, increases as the lag decreases.  We compute
below the largest statistically significant lag in each available map. We
request that the largest lag provides at least a minimum number of 10
independent elements in each bin of the \Cincr{}--PDFs.

We call \Nsign{} the minimum number of {\it independent} measurements
required to build a statistically meaningful PDF. From the above condition,
\Nsign{} is at least 10 times the number of bins $N_\emr{bin}$ in the PDF.
The PDFs shown in Figs.~\ref{fig:PDFs:pol} and~\ref{fig:PDFs:l1512} are
built with 30 bins therefore $\Nsign \approx 300$.

To compute \Nsign{}, we consider that two increments are independent if
\emph{i)} the two pairs of points have at most one point in common and
\emph{ii)} the length between the two points which differ is at least equal
to the lag $l$ used to compute the increments. This makes the reasonable
assumption that in the inertial range of turbulence, variations of the
velocity at scale $l$ decorrelate over the distance $l$.  We therefore
divide the map in cells of size $l^2$. In each cell, we find four
independent pairs of points: the two diagonals and half the four cell edges
(since each edge belongs to two contiguous cells).  Hence, there are about
$4\xsize\ysize/l^2$ independent increments in a map of $\xsize \times
\ysize$ pixels. Thus,
\begin{equation}  
\Nsign{} \about 4 \frac{\xsize\ysize}{l^2}
\end{equation}
from which we derive the largest lag, $l_\emr{max}$, for which the PDF is
statistically significant, as a function of the number of bins in the PDF:
 \begin{equation}
  l_\emr{max}(N_\emr{bin}) \about \sqrt{\frac{2}{5} \frac{\xsize\ysize}{N_\emr{bin}}}.
\end{equation}
Table~\ref{tab:lags} gives numerical values of the largest reliable lag for
the two cases of \Cincr{}--PDFs built with 30 and 8 bins.  We see, as
expected, that the most statistically significant PDFs are those obtained
for the large scale field of L1512, and that at least for the two smallest
lags (\ie\ 3 and 6 pixels), the PDFs obtained with the other fields have
the same significance.  Table~\ref{tab:lags} shows that all the PDFs of
Figs.~\ref{fig:PDFs:pol} and~\ref{fig:PDFs:l1512} computed for $l=3$ and 6
pixels are statistically significant as drawn (\ie{} with 30 bins). Those
computed for lags $l = 12$ pixels in the IRAM-30m maps are significant only
for their smooth characteristics, \ie\ those visible if the PDFs were built
with 8 bins only, as shown with the column $l_\emr{max}(8)$.

\end{document}